\documentstyle[11pt,aaspp4]{article}

    \newcommand{\Kp}{$K^\prime$}
    \newcommand{\R}{$R$}
    \newcommand{\mgn}{$^{\rm m}$}

\slugcomment{{\em to appear in ApJ, September 1999}}

\lefthead{Thompson et al.}
\righthead{Surface Density of Extremely Red Objects}

\begin{document}

\title{The Surface Density of Extremely Red Objects\\}

\author{D. Thompson\footnote{Current address: MS 320-47 Caltech, Pasadena, 
        CA  91125, USA.  Email: djt@mop.caltech.edu}, S. V. W. 
        Beckwith\footnote{Current address: STScI, 3700 San Martin Dr., 
        Baltimore, Md. 21218 USA.  Email: svwb@stsci.edu}, R. Fockenbrock, 
        J. Fried,\\
        H. Hippelein, J.-S. Huang, B. von Kuhlmann, Ch. Leinert,\\
        K. Meisenheimer, S. Phleps, H.-J. R\"oser, E. 
        Thommes\footnote{Current address: Royal Observatory, Blackford Hill, 
        Edinburgh, EH9 3HJ, UK.  Email: emt@roe.ac.uk}, C. Wolf}
\affil{Max-Planck-Institut f\"ur Astronomie,\\ 
       K\"onigstuhl 17, D-69117 Heidelberg, Germany\\
       Electronic mail: djt,svwb,fock,fried,hippelei,huang,kuhlmann,leinert,\\
       meise,phleps,roeser,thommes,cwolf@mpia-hd.mpg.de}

\begin{abstract}
We present initial results from a field survey for extremely red
objects (EROs, defined here as (\R-\Kp) $\geq$ 6\mgn) covering 154
square arcminutes of sky, from the first of 7 deep, wide-field
\Kp\ images obtained as part of the Calar Alto Deep Imaging Survey
(CADIS).  The 5$\sigma$ point source detection limits are \Kp\ $=$
20\fm5 and \R\ $=$ 25\fm0, while extended-source limits are up to 
0\fm50--0\fm75 brighter.  We identify a total of 8 bright EROs with
\Kp\ $\leq$ 19\fm0.  Six of these bright EROs are resolved and are
likely to be galaxies, while the remaining 2 are unresolved, with 
colors consistent with their being low-mass Galactic stars.  We 
derive a surface density for the 6 bright, extragalactic EROs of 
0.039$\pm$0.016 arcmin$^{-2}$, higher by a factor of 4 than previous 
values.  We estimate that the volume density of bright EROs to be as 
high as that of nearby Seyfert galaxies.
\end{abstract}


\clearpage

\section{Introduction}

Significant numbers of massive galaxies have been identified 
in the near-infrared with colors so red that they are not 
found in surveys that select galaxies at visual wavelengths.  Such 
objects have been named {\em extremely red objects} (EROs).  With 
surface densities similar to that of bright quasars (Hu \& Ridgeway, 
1994), they represent an important component of the population of 
high-redshift galaxies.  

The identification of EROs coincided with the development and
implementation of infrared arrays for astronomical research, the first
EROs being noted in the K-band surveys of Elston, Rieke, \& 
Rieke (1988, 1989).  Since then, many groups have identified objects 
with extremely red colors (McCarthy, Persson, \& West 1992; Eisenhardt 
\& Dickinson 1992; Persson et al. 1993, Hu \& Ridgeway 1994, Soifer et al. 
1994, Dey, Spinrad \& Dickinson 1995, Djorgovski, et al. 1995, among 
others), though often with widely different definitions of {\em extremely 
red}.  These objects were primarily identified in the fields around 
high-redshift active galactic nuclei, i.e. around radio galaxies and 
quasars.  In this paper, we define the selection criteria for bright EROs 
to be \Kp\ $\leq$ 19\fm0 and with (\R-\Kp) $\geq$ 6\mgn.

The broad-band spectral energy distributions (SEDs) of these galaxies 
were initially thought to be fitted best by an old stellar population, 
such as found in present-day elliptical galaxies.  In this scenario, a 
redshifted ($z > 0.85$) strong 4000\AA-break falling between the \R\ and 
$K$ filter bandpasses and the lack of any appreciable restframe UV light 
from a young population of stars are responsible for the extremely 
red colors.  An alternative interpretation is that these galaxies are 
starbursts or active galactic nuclei (AGN), perhaps triggered by a 
merging event.  In this scenario, the presence of significant quantities 
of interstellar dust hides the star-forming regions or AGN, considerably 
reddening the observed SEDs.  These would also lie at redshifts $1 < z 
< 2$, which shifts the restframe UV into the optical bandpass while 
still sampling the restframe optical/near-IR wavelengths in the K filter.  

Archetypal examples of each of these interpretations have been 
identified.  LBDS 53W091 (Dunlop et al. 1996) is a 
faint radio galaxy at a redshift of 1.55, derived from an absorption 
line spectrum obtained at restframe UV wavelengths.  Spectral 
modeling suggests an elliptical galaxy spectrum with an age of up to 
3.5 Gyr (Dunlop et al. 1996, Spinrad et al. 1997, but see Bruzual \& 
Magris 1997 for an alternative viewpoint), implying a much higher 
redshift of formation for this system.  In contrast, HR10 (Hu \& 
Ridgeway 1994) has been identified as a dusty starforming galaxy, 
perhaps with an active nucleus, at $z=1.44$ (Graham \& Dey 1996).  
Detections of this galaxy at submillimeter and millimeter wavelengths 
support the interpretation that the SED is dominated by the presence 
of dust (Cimatti et al. 1998).

In both cases, the EROs are most likely to lie in the redshift range $1
< z < 2$.  It is difficult to produce such red colors at lower
redshifts, while higher redshift objects become exceedingly luminous.
If EROs are dominated by old stellar populations, then massive galaxy
formation was well underway at $z > 3$.  Dust-dominated EROs, however,
imply that much of the massive galaxy formation could actually occur at
late times, supporting hierarchical galaxy formation models.  The
contrasting implications these two scenarios have for the history of
galaxy formation is the primary driver behind continuing work on these
objects.

We have therefore started a wide-area field survey to select 
an unbiased and statistically significant sample of EROs for further 
study.  This paper presents initial results from the first of seven 
fields, obtained as part of the Calar Alto Deep Imaging Survey (CADIS, 
Meisenheimer et al. 1997), each of which will cover $\sim$160 square 
arcminutes.  Section 2 describes the observations and reductions.  
Section 3 includes a discussion of the sample selection, morphology, 
and surface and volume density of the EROs identified here.  Section 
4 gives a brief summary of this paper.

  
\section{Observations and Reductions}

The field observed in this paper is centered on $\alpha_{1950}$ $=$ 
16$^h$ 23$^m$ 29$^s$ $\delta_{1950}$ $=$ $+55^\circ$ 50$^\prime$ 
47$^{\prime\prime}$ (hereafter the 16\,h field).  It is one of 7 fields 
in the CADIS survey (Meisenheimer et al. 1997), chosen to be free of 
bright stars or galaxies, known galaxy clusters, and in regions of 
minimal extinction due to Galactic cirrus.  All magnitudes quoted 
in this paper are referred to $\alpha$ Lyra.

\subsection{K$^\prime$ imaging}

The \Kp\ imaging was obtained with the Omega-Prime camera (Bizenberger 
et al. 1998) at the prime focus of the Calar Alto 3.5m telescope on UT 
1996 June 7-8 and UT 1996 August 22-29.  The majority of these nights 
were photometric.  Omega-Prime is a direct imaging camera equipped with 
a HAWAII 1024$^2$ HgCdTe array.  The image scale is 0.396"/pixel (6.75 
arcminute square field of view).  Data were taken in a 2$\times$2 mosaic 
to cover as much of the CADIS survey field (16$^\prime$ diameter) as 
possible.  The \Kp\ filter ($\lambda_{\rm cent} = 2.12$\,$\mu$m, $\Delta 
\lambda = 0.35$\,$\mu$m, see Wainscoat and Cowie 1991) was used because 
its bluer bandpass relative to a standard K filter significantly reduces 
the thermal background seen by the array.

The data were reduced with a standard infrared reduction algorithm in 
IRAF\footnote{IRAF is distributed by the National Optical Astronomy 
Observatories, which are operated by the Association of Universities 
for Research in Astronomy, Inc., under cooperative agreement with the 
National Science Foundation.}.  A sky frame was constructed from 6 to 
12 temporally adjacent images, scaled to have the same median counts 
before averaging all but the 1 or 2 most extreme values in each pixel, 
effectively removing stars from the final sky image.  The sky frame was 
then scaled to and subtracted from the data frame and the result divided 
by a normalized dome flat to remove pixel-to-pixel fluctuations in 
quantum efficiency.

On photometric nights, standard stars from the UKIRT faint standards
list (Casali \& Hawarden 1992) were observed every 2-3 hours at several
different airmasses to determine both a photometric zero point and an
extinction correction.  Individual images were then corrected for
extinction and put onto a common photometric scale.  Night to night
variations of the zero points were under 0.1 magnitude, giving a
measure of the systematic uncertainty in the \Kp\ calibrations.

As IRAF was not able to handle either the large number of frames (1209)
or the large size of the data set (4.8\,GB), a procedure for IDL was
developed to stack the individual images into the final mosaic.  Known
bad pixels were ignored and only good pixels were included in the
mosaic.  The software used a multi-pass sigma-clipping algorithm to
reject cosmic rays and other transient phenomena (e.g meteors and
satellites).

\begin{figure}
   \epsscale{0.5}
   \plotone{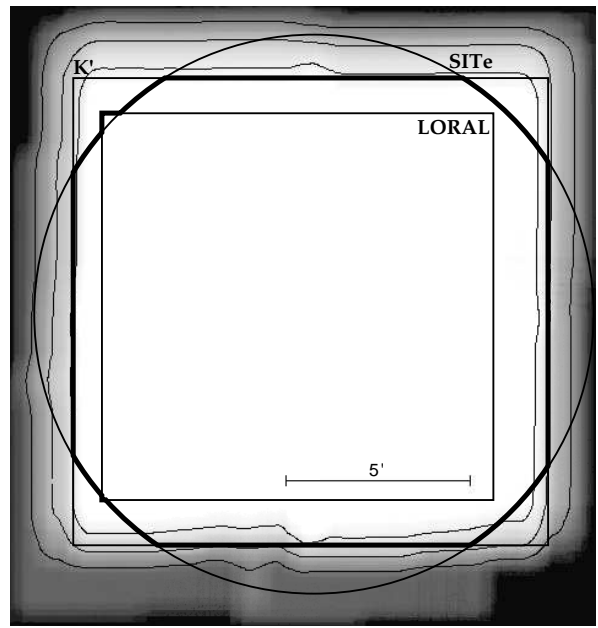}
   \figcaption[EROs_C16.fig1.eps]{The grayscale background image shows the 
               relative exposure time in the full \Kp\ mosaic.  Three contours, 
               at 1000, 3000, and 6000 seconds exposure time, are also plotted. 
               The area used for the ERO survey is indicated (heavy line), 
               showing the area of overlap between the K' image (outer 
               square) and the two \R\ images (LORAL (inner square) and SITe 
               (circle) CCDs).  
               \label{xtmap}}
   \epsscale{1.0}
\end{figure}

The final mosaic is 2419 $\times$ 2546 pixels (15\farcm96 $\times$
16\farcm80), including regions of higher noise around the edge where
coverage was incomplete.  The deepest portion of the mosaic, with per
pixel exposure times greater than 6000 seconds, covers the central
$\sim$12.5 arcmin square with a seeing of 1.1 arcsec full-width at
half-maximum (FWHM).  A grayscale image of the exposure time map, along
with the relative fields of view for the optical and infrared data, is
shown in Figure~\ref{xtmap}.  The central portion of the \Kp\ mosaic
has a 5$\sigma$ point source detection limit of \Kp\ $=$ 20\fm5, within
an aperture diameter of 2\farcs2, twice the seeing FWHM.  Limits for 
extended sources are up to 0\fm50--0\fm75 brighter, depending on the 
object morphology.

\placefigure{xtmap}	

\subsection{\R-band imaging}

The \R-band imaging was obtained with the CAFOS reimaging camera at the
Calar Alto 2.2\,m telescope on UT 1996 May 15 (5$\times$500\,s,
1\farcs3 seeing) and UT 1997 February 5 (4$\times$700\,s 1\farcs6
seeing).  CAFOS was equipped with a LORAL 2048$^2$ CCD
(0.33$^{\prime\prime}$/15\,$\mu$m pixel) for the May observations, and
with a SITe 2048$^2$ CCD (0.53$^{\prime\prime}$/24\,$\mu$m pixel
covering a circular field of view $\sim$16 arcminutes in diameter) for
the remaining observations.  The \R\ filter used in CADIS ($\lambda_{\rm
cent} = 648$\,nm, $\Delta \lambda = 171$\,nm) is slightly narrower and
bluer than a standard Johnson \R\ filter.  This difference has no 
significant impact on the goals of this survey.

The data were reduced with the MIDAS software package.  Following bias
subtraction and flatfielding, multiple exposures obtained on a single
night were shifted into coincidence, bad pixels and cosmic rays masked
out, then summed to form a deep image for that night.

Flux calibration of the \R-band data was established relative to two
stars in the 16\,h field for which we have good spectrophotometry.  These
stars were calibrated against an HST UV spectrophotometric standard
(AGK$+81^\circ266$) as well as 3 stars from Oke (1990).  The magnitudes
of the two 16\,h field reference stars in the \R\ filter were determined by
integrating their calibrated spectra over the \R\ filter bandpass and
instrumental transmission curve.  The 2$\sigma$ limits were established
within an aperture diameter twice the seeing FWHM, scaling from the
per-pixel noise in the stacked data.  A full description of the
calibration of the CADIS optical data set can be found in Fockenbrock
(1998).

\subsection{Object Detection}

Objects were separately identified on the deep \R\ and \Kp\ images using
the Source Extractor software (Bertin \& Arnouts 1996) and the
resulting lists merged into a master catalog.   This catalog was then
used to extract photometric magnitudes for all objects in each
individual exposure (mosaics from each night for the \Kp\ data) using
photometry software (Meisenheimer \& R\"oser 1986) incorporating a
Gaussian weighting function, designed to compensate for variations in
seeing between individual exposures.  The procedure recovers accurate
{\em colors} for all objects, but underestimates the total magnitude
for extended objects.  For the faint galaxy images from which the
sample of EROs was drawn, typically subtending only
1-2$^{\prime\prime}$, the underestimate is of order 0.1 magnitude.

A final list of objects was selected from the master catalog,
containing objects with \Kp\ magnitudes above the 5$\sigma$ limit of
20\fm5 that also appear on at least one of the two \R\ images (from the
LORAL and SITe CCDs).  Figure~\ref{xtmap} shows a contour map and image
of the exposure time, overlaid with the outline of the survey area.
The full area imaged in both \Kp\ and \R\ covers 154 square arcminutes.

\section{Discussion}

\subsection{ERO Selection}

\begin{figure}
   \plotone{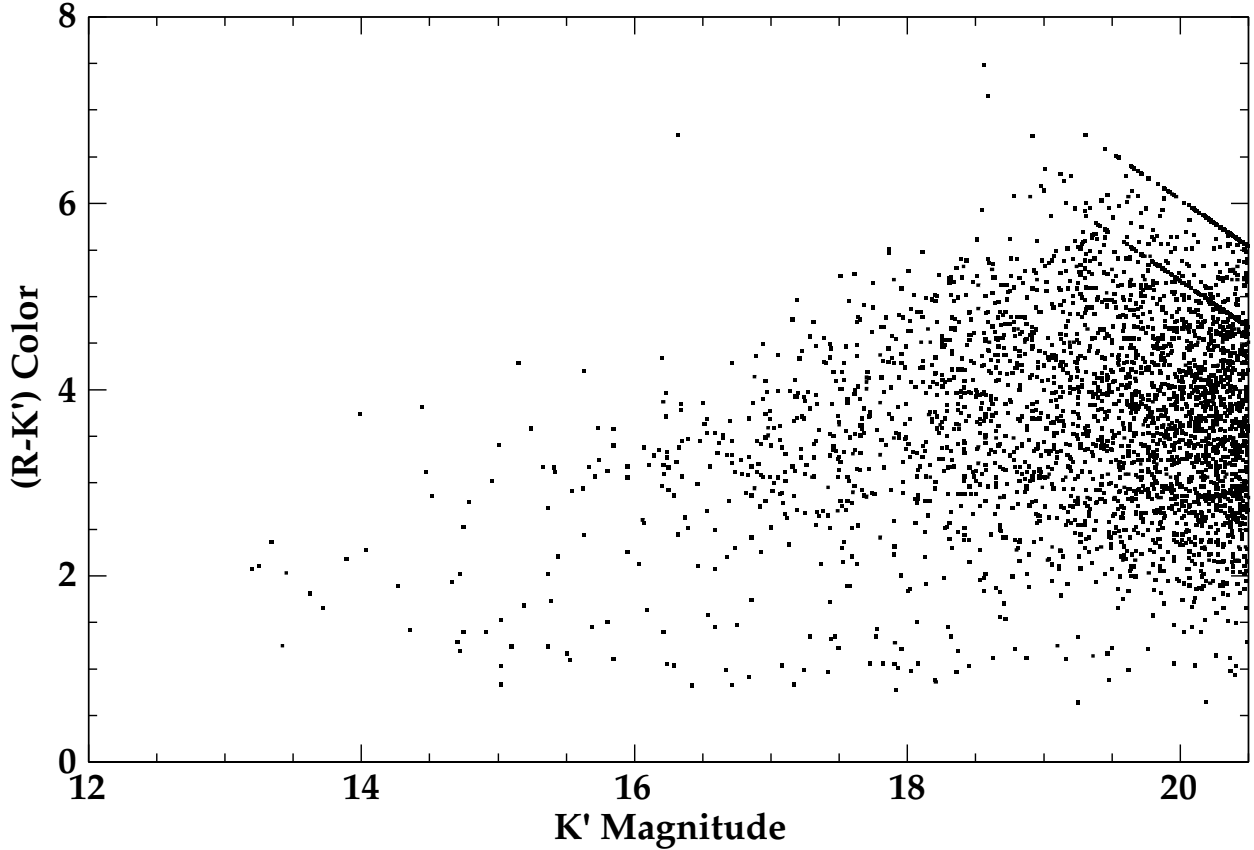}
   \figcaption[EROs_C16.fig2.eps]{The \R-\Kp\ color-magnitude diagram for the 
               entire CADIS 16\,h field.  Two loci of points on diagonal 
               lines are visible in the upper right corner.  These loci are 
               artifacts resulting from the use of different sized \R\  
               images (see section 2.2 and Figure~\ref{xtmap}).  The deeper  
               limit represents the effective 2$\sigma$ limit of 26\fm04  
               reached in the combined \R\ images, while the brighter limit 
               corresponds to the 2$\sigma$ limit of 25\fm17 in the \R\  
               image taken with the SITe CCD but outside of the area covered 
               by the LORAL CCD.  Objects which are undetected in the \R\  
               data are plotted at the 2$\sigma$ limit.  The majority of 
               objects with \R-\Kp\ $<$ 2 are stars.
               \label{fullcmd}}
   \epsscale{1.0}
\end{figure}

\begin{figure}
   \epsscale{0.5}
   \plotone{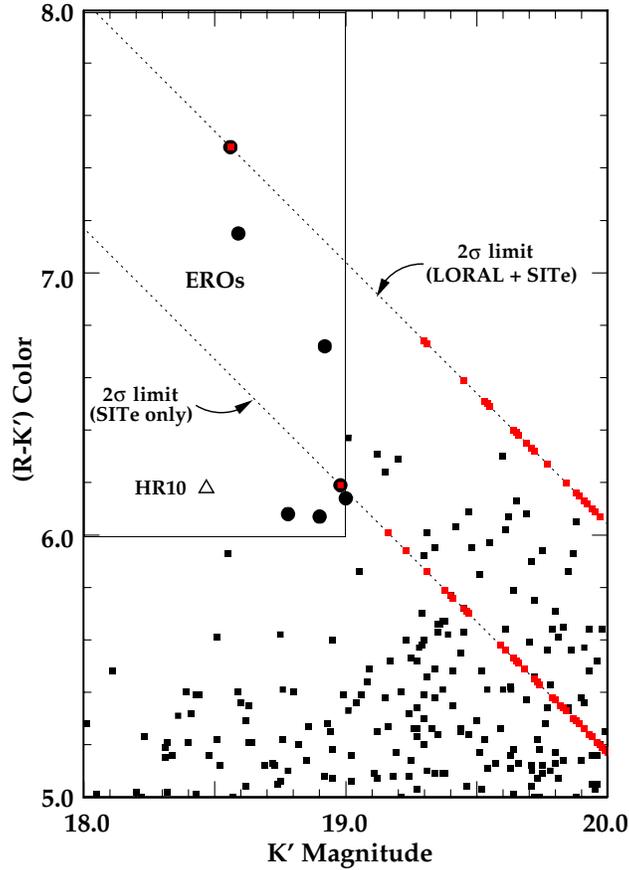}
   \figcaption[EROs_C16.fig3.eps]{Enlarged portion of Figure~\ref{fullcmd} 
               covering the area near the ERO selection region.  Seven 
               of the 8 bright EROs are marked with the large filled 
               circles, the last ERO is off the plot to the left 
               (at \Kp\ $=$ 16\fm3).  The remaining objects from the 
               16$^{\rm h}$ field are plotted as filled squares.  Objects 
               which fall on the two diagonal lines are undetected in the 
               \R\ data at the 2$\sigma$ level (see Figure~\ref{fullcmd} 
               caption for details) and should be considered lower limits 
               to the actual color.  The location of HR 10 (open triangle) 
               is shown for comparison.  No published \R-band magnitude 
               for HR10 exists.  We estimate the \R\ magnitude to be 
               \R$\sim$25\fm0, derived from a linear interpolation between 
               the B and I band fluxes for HR10 from Graham \& Dey (1996).  
               \label{zoomcmd}}
   \epsscale{1.0}
\end{figure}

\begin{figure}
   \epsscale{0.5}
   \plotone{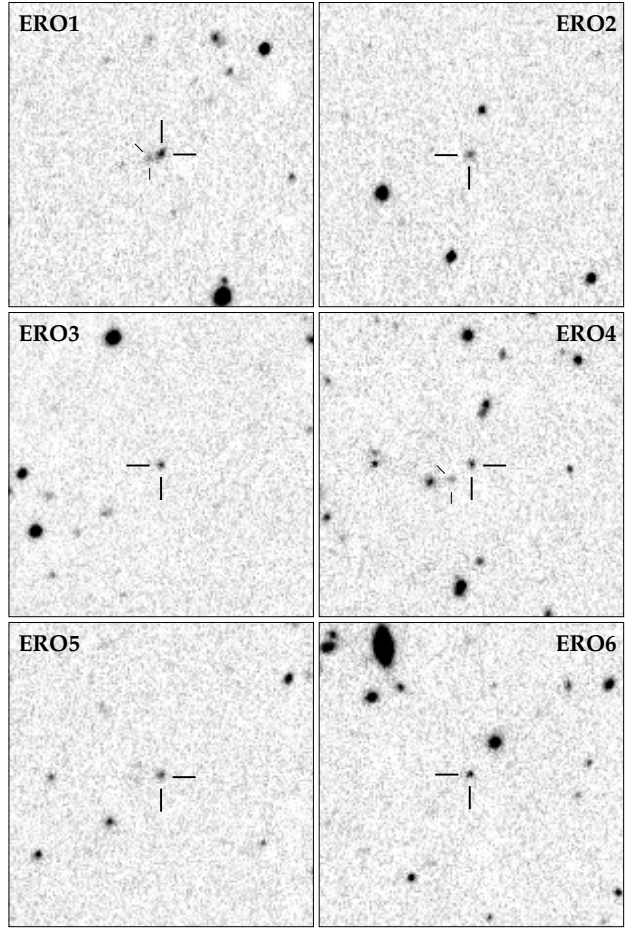}
   \figcaption[EROs_C16.fig4.eps]{Finding charts for the 6 bright EROs.  Each 
               subimage is 60$^{\prime\prime}$ square, with north up and 
               east to the left.  Both ERO1 and ERO4 have red companion 
               galaxies to the southeast, indicated with the smaller tick 
               marks.    
               \label{FCs}}
   \epsscale{1.0}
\end{figure}

We select objects with extremely red colors from the color-magnitude 
diagram for the full survey area, shown in Figure~\ref{fullcmd}.  We 
define the selection criteria for bright EROs as objects with \Kp\ 
$\leq$ 19\fm0 and (\R-\Kp) $\geq$ 6\mgn.  A more detailed plot of 
this region of the color-magnitude diagram is shown in Figure~\ref{zoomcmd}. 
Eight objects in our survey field meet these selection 
criteria.  Their positions, magnitudes and colors are summarized in 
Table~\ref{ERO_data}, while finding charts from the \Kp\ data are 
given in Figure~\ref{FCs}.    

\placefigure{fullcmd}	

\placefigure{zoomcmd}	

\placetable{ERO_data} 

\placefigure{FCs}   

The magnitude limit defining the bright sample of EROs is partially driven 
by the current data: most EROs brighter than \Kp\ $\leq$ 19\fm0 are 
detected above the 2$\sigma$ limit in \R\, while most of those fainter 
than this are not.  This also represents the practical limit for obtaining 
followup near-infrared spectroscopy at 8m-class telescopes (Moorwood \& 
Spyromilio 1997) with exposure times of order 1 hour.  

In the absence of significant dust extinction, the reddest galaxy colors 
will come from the oldest stellar populations at any redshift.  A 
present-day elliptical galaxy SED under a no-evolution model peaks in 
color at $1 \lesssim z \lesssim 3$ (ignoring the Lyman break at higher 
redshift), with an \R-\Kp color of about 6.  For an old stellar population, 
this color is produced by a combination of the cosmological k-corrections 
and a strong 4000\AA-break lying between the two bandpasses.  Since all 
galaxies should be bluer than this limit, selecting objects with \R-\Kp 
$\geq$ 6$^{\rm m}$ will isolate a sample of galaxies with extremely red 
colors.  The specific choice of \R-\Kp $\geq$ 6\fm0 is, however, also 
motivated by historical precedence, approximating the limits used to 
identify other EROs in the literature.  Such objects were identified as 
lying away from the locus of other field objects in the color-magnitude 
diagram.  However, the areas surveyed, and thus the number of objects 
detected, were generally too small to address this point statistically.   

\begin{figure}
   \epsscale{0.5}
   \plotone{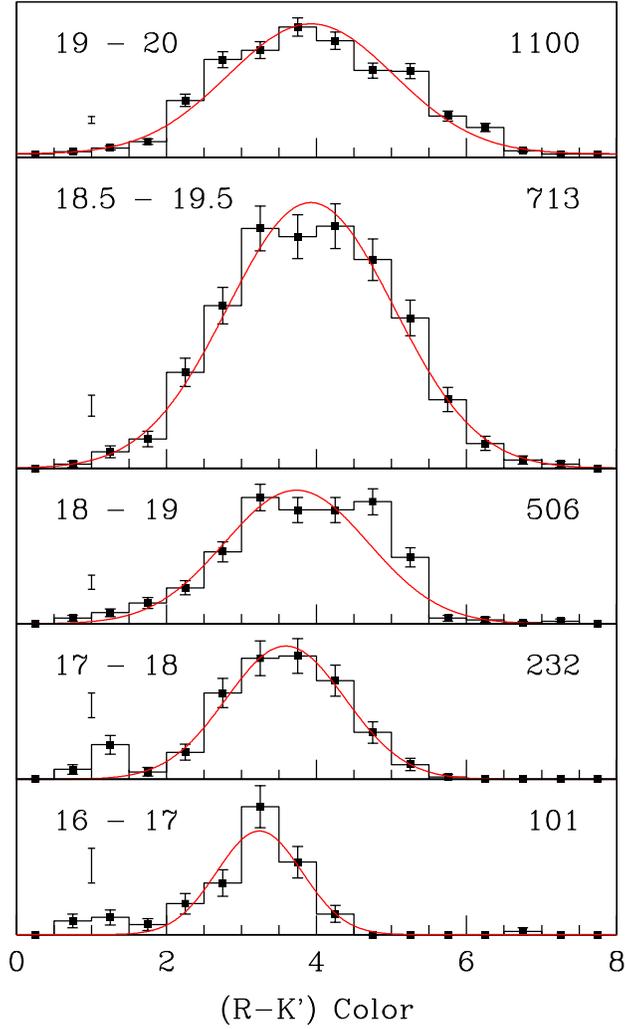}
   \figcaption[EROs_C16.fig5.eps]{Color histograms for different magnitude 
               bins in the 16\,h field data.  Magnitude ranges are 
               indicated in the upper left corner of each graph, the 
               total number of objects used for each subpanel in the 
               upper right corner.  The histograms show the actual data, 
               with poisson uncertainties indicated.  Mean color and FWHM 
               for the Gaussian fits are given in Table~\ref{colorhist_data}.  
               The vertical scale bars at color $=$ 1.0 in each subpanel 
               are 10 units high.  
               \label{colorhist}}
   \epsscale{1.0}
\end{figure}

Figure~\ref{colorhist} displays histograms of the \R-\Kp colors for objects 
with 16\mgn\ $\leq$ \Kp\ $\leq$ 20\mgn\ in 1\mgn\ bins.  
Table~\ref{colorhist_data} lists the mean color and FWHM for the 
Gaussian least-square fits which are also plotted in Figure~\ref{colorhist}.  
For example, in the range 18\fm5 $\leq$ \Kp\ $\leq$ 19\fm5, the bin containing 
most of the EROs, the mean color is 3\fm92 and the standard deviation is 
1\fm11.  An (\R-\Kp) $\geq$ 6\mgn selection criterion therefore corresponds 
to colors that deviate by more than about two standard deviations from the 
mean color.  The extremely red galaxies selected here are thus the reddest 
2\% of the population and should consist of the oldest and/or the dustiest 
high redshift galaxies.

\placefigure{colorhist}	

\placetable{colorhist_data} 

There is a distinct trend to redder colors at fainter magnitudes, with 
the mean (\R-\Kp) color becoming 0\fm7 redder between \Kp\ $=$ 
16\fm5 and \Kp\ $=$ 19\fm5.  This continues the trend noted at 
brighter magnitudes in (I-K) colors by Huang et al. (1997).  Over 
the range sampled by our data, this trend probably reflects an increasing 
fraction of high redshift galaxies.  

There are an additional 43 objects with (\R-\Kp) $\geq$ 6\mgn in the 
range 19\fm0 $<$ \Kp\ $\leq$ 20\fm0.  These fainter EROs lie only in 
the central 120 square arcminutes of the survey field, where the 
overlap of the two sets of \R-band data reach deep enough to select 
EROs with \Kp\ $\leq$ 20\fm0.  We note that since both the color and 
color dispersion for the objects in the 16$^{\rm h}$ field increase 
with increasing magnitudes, the EROs in this fainter magnitude range 
may no longer have truly extreme colors.  The current optical data do 
not have sufficient depth to address this more quantitatively.  

We chose to set the detection limit in the \R\ data at 2$\sigma$ to 
ensure that all of the EROs are selected, at the expense of some possible 
contamination of the sample.  In practice, contamination is likely to be 
low.  Consider a Malmquist bias vector defined on the color-magnitude 
diagram (Figure~\ref{zoomcmd}).  For objects near the ERO selection 
region, it would be dominated by the uncertainties in the \R-band 
magnitudes, since most of the objects there are well-detected at \Kp\ 
but near the 2$\sigma$ detection limit in \R.  The Malmquist bias vector 
would thus point to {\em bluer} colors and slightly brighter \Kp\ 
magnitudes.  Inspection of Figure~\ref{zoomcmd} shows that there are 
a comparable number of objects which could scatter into the ERO region 
as are already there, so contamination of the sample is likely to be 
low and would be coming from objects with redder colors.

\subsection{Morphology and Environment}

The radial brightness profiles of the EROs were compared to those of
several stars near each object to determine whether the EROs are
resolved.  Five of the 8 bright EROs are easily resolved, with FWHMs 
ranging from 1\farcs4 to 2\farcs3 in 1\farcs1 seeing, and are therefore 
likely to be galaxies.  Subtracting the intrinsic seeing in quadrature 
gives deconvolved FWHMs ranging from 0\farcs9 to 2\farcs0.  Such sizes 
are similar to both HR10 (0\farcs7, Hu \& Ridgeway 1994) and 53W091 
(1\farcs3, Dunlop et al. 1996).  A sixth object, C16-ERO3, is formally 
resolved, but at a low statistical significance, with a deconvolved 
FWHM $\leq$ 0\farcs4.

The remaining 2 bright EROs in the 16\,h field are unresolved.  Their 
broadband colors indicate that they are low mass Galactic stars.  One 
has subsequently been spectroscopically confirmed as a low mass star of 
spectral type L1 (Wolf et al. 1998).  The second is well-detected in 
the CADIS filters (\Kp $=$ 16\fm32), with colors fully consistent with 
a star of spectral type M9 (Wolf et al. 1998), though without spectroscopic 
confirmation we cannot completely rule out that the object may be a compact 
galaxy or AGN at high redshift.  Finding charts for these two objects 
can be found in Wolf et al. (1998), and they are not considered further 
in this paper.

Only one of the six resolved EROs is significantly non-circular (C16-ERO1).  
The deconvolved FWHMs along the major and minor axes are 1\farcs9 and 
1\farcs1, respectively, with the major axis oriented at a position angle 
(PA) of 143$^\circ$, east of north.  There is also a companion ERO with 
\Kp\ $>$ 19\fm0, 2\farcs8 away at PA $=$ 126$^\circ$.  The proximity of 
the second ERO and the apparent alignment between the PA of C16-ERO1 and 
the direction to the fainter galaxy suggest that this might be an 
interacting system.  C16-ERO1 and C16-ERO2 also make the closest pair 
of EROs in the current sample, separated by 39\farcs2 at PA $=$ 125\fdg5, 
possibly indicating the presence of a cluster or group.  One other object 
(C16-ERO4) has a fainter ERO companion nearby (4\farcs9 away at PA $=$ 
127$^\circ$), though there is no evidence for interaction in the current 
data set.

Of the fainter 43 EROs, 37 are either clearly or marginally resolved.
Below K$^\prime \sim$ 19\fm5, it becomes difficult to distinguish
between unresolved and slightly extended sources on the survey data.
Nevertheless, the fainter EROs in this field are clearly dominated by
galaxies.

\subsection{Surface Density}

The 6 bright, extragalactic EROs identified here were found in an area
of 154\,arcmin$^2$, yielding a surface density of 0.039 $\pm$ 0.016
arcmin$^{-2}$ (poisson uncertainties).  This density is a factor of
four higher than found by Hu \& Ridgeway (1994), estimated from 
100\,arcmin$^2$ of I and \Kp\ data from the Hawaii Deep K-band Surveys
(Cowie et al. 1994).  In their survey, Hu \& Ridgeway found no galaxies 
as red as HR10, implying a surface density of $\lesssim$\,0.01\,arcmin$^{-2}$.  
The difference is not statistically significant due to the currently large 
uncertainties from the small number statistics as well as a possible 
contribution from field-to-field variations.  Combining these data with 
those from future fields obtained as part of this survey will reduce this 
uncertainty to $\sim$15\%.  Still, the surface density of EROs is 
comparable to or higher than that of bright ($B < 22$\mgn) quasars, making 
the EROs a significant population.

Dey, Spinrad, and Dickinson (1995) suggest that EROs are more common 
along the line of sight to or clustered with high redshift radio 
galaxies and quasars.  Estimates of the surface density range up to 
10-100 times that of the field EROs.  If the EROs are at the same 
redshifts as the high-redshift AGN, then many of them would have 
luminosities significantly in excess of $L^*$.  With the EROs in the 
foreground there is the possibility of a magnification bias, where the 
ERO (or a group or cluster containing it) is acting as a gravitational 
lens for the background AGN.  This higher surface density in the fields 
of high-redshift AGN may also reflect what we term an ``observational 
bias.''  This is a combination of ({\em i}) AGN were a common target for 
observations in the near-IR, ({\em ii}) unusual objects of any sort were
noted (here objects with extremely red colors) when the data were published, 
and ({\em iii}) earlier generations of near-IR imagers had fields of 
$\sim$\,1\,arcmin$^2$.  A systematic measurement of the surface density 
in AGN fields still remains to be done.  

Guiderdoni et al. (1997, their model C) predict the surface density
($\sim 0.05$\,arcmin$^{-2})$ of high-redshift dusty star-forming
galaxies observed at far-infrared wavelengths that is remarkably
similar to the surface density of bright EROs found here, suggesting
that EROs may be dusty starbursts.  The predictions were based on local
ultra-luminous infrared galaxies, actively forming stars but completely
shrouded by dust.  The detection of significant quantities of dust in
HR 10 (Cimatti et al. 1998) supports this picture.

Including the additional 43 EROs with \Kp\ $\leq$ 20\mgn, the 
surface density is 0.33$\pm$0.05 arcmin$^{-2}$.  As noted above, 
the fainter sample is not well defined over the survey area, and 
this is only an estimate to the true surface density of faint EROs.  
The depth of the \R\ band data is currently the limiting factor, and 
deeper imaging will allow us to make a stronger statement about 
these fainter EROs.

\subsection{Volume Density}

We can derive a volume density of EROs if we can estimate the 
redshift range over which they occur.  This volume density would 
then be directly comparable to known populations of galaxies, perhaps 
providing some insight into their nature.

If the ERO population is dominated by old (elliptical) galaxies, the 
red color would be coming from a strong 4000\AA\ break.  This spectral 
break comes primarily from Balmer continuum limit and Ca H$+$K 
absorption lines and can significantly redden an (\R-\Kp) color when 
the break lies between the two filter bandpasses.  For the \R\ filter 
used here, this implies a redshift beyond $z=0.85$.  

If, on the other hand, the ERO population is dominated by dusty
starburst galaxies, the red color would be due to dust extinction
reddening a flat SED.  A color of (\R-\Kp) $\sim$ 6\mgn\ can be produced
at $z = 0.85$ by a V-band extinction of $A_V \sim 5^{\rm m}$, assuming
the Seaton (1979) Milky Way extinction law.  In the grayer extinction
law of Calzetti et al. (1994), derived empirically from a sample of
local starburst galaxies, a factor of two more extinction would be
required to give similar results.  The steeply rising extinction curves
at shorter wavelengths, however, imply that less dust extinction would
be required to produce the same reddening at higher redshifts.  In
practice, a moderate continuum decrement across the 4000\AA\ break is
also seen in relatively young stellar populations, contributing to the
red color and relaxing the extinction requirements.  We therefore adopt
a redshift of $z=0.85$ as the lower redshift limit for both types of
galaxies.

For the high redshift limit, consider an L$^*$ galaxy at $z = 2$.  It 
would have an apparent magnitude of \Kp\ $\simeq$ 20\fm3 (H$_0 = 
70$\,km\,s$^{-1}$\,Mpc$^{-1}$ and q$_0 = 0.1$), assuming only passive 
evolution (evolution and k-corrections from Poggianti 1997 and Fioc \& 
Rocca-Volmerange 1997 give similar results).  Our selection 
of bright EROs at \Kp\ $\leq$ 19\mgn\ is thus already sampling 
the luminosity function well above L$^*$ at $z = 2$.  Beyond this 
redshift, the galaxies would have to be intrinsically as luminous as 
bright quasars to be selected as EROs.  We therefore adopt the range 
$z \in [0.85,2.0]$ where EROs are most likely to be found with the 
selection criteria used here.  

The comoving volume covered in this redshift range over our survey 
field in the assumed cosmology is 3.55$\times10^5$\,Mpc$^3$.  For 
the six bright EROs in our sample, this yields a volume density of 
$1.7 \pm 0.7 \times 10^{-5}$\,Mpc$^{-3}$.  Extending the redshift 
range down to $z = 0.5$ would only increase the comoving volume by 
15\%.  Increasing the upper redshift limit to $z = 3$ would 
approximately double the sampled volume.  The volume density derived 
here is, therefore, not very sensitive to the assumed redshift range.  

We can compare this estimate of the volume density of EROs with that 
of other known types of objects.  For quasars (M$_{\rm B \leq -23^m}$), 
the volume density in a similar redshift range (Boyle 1991) is only $2 
\times 10^{-6}$\,Mpc$^{-3}$, a factor of 8 lower than the bright ERO 
density.  A simple extrapolation of the Boyle luminosity function to 
M$_{\rm B} \sim -19^m$ (it is strictly only valid for M$_{\rm B} \leq 
-21^m$) is required to reach space densities similar to those of the 
EROs.  Thus the {\em bright} ERO galaxies appear to be considerably 
more numerous than quasars, and perhaps as common as Seyfert Is in 
the local universe.  The EROs, however, have been compared to both 
old ellipticals and young dusty starbursts, and it is perhaps more 
interesting to compare them to these groups of objects. 

The red galaxies from the Canada-France Redshift Survey (CFRS, Lilly 
et al. 1995) represent a population of objects which to not undergo 
significant evolution from $z \sim 1$ to the present, similar to what 
would be expected of the EROs if they are dominated by old elliptical 
galaxies.  The integrated luminosity function for the CFRS red galaxies 
reaches a volume density comparable to the EROs at a luminosity of 
$\sim$4\,L$^*$.  Thus, if the ERO population is dominated by old 
elliptical galaxies, then the EROs are very massive and suggest that 
the most massive systems collapsed early in the history of the universe.  
Such galaxies at $z \sim 1.5$ could also mark the location of 
high redshift (forming?) clusters or groups of galaxies.

Luminous infrared galaxies are often powered by strong starbursts or 
AGN, but emit most of their bolometric luminosity at restframe far-infrared 
wavelengths due to the presence of significant quantities of dust.  The 
ERO population could be high-redshift analogs to these luminous infrared 
galaxies if the EROs are dominated by objects like HR10.  The integrated 
luminosity function for the luminous infrared galaxies (Soifer \& 
Neugebauer 1991) reaches a volume density comparable to the EROs at a 
bolometric luminosity of a few\,$\times\,10^{11}L_\odot$, corresponding 
to $\sim$10\,L$^*$.  Thus, if the ERO population is dominated by young, 
dusty starbursts similar to the luminous infrared galaxies, then many 
massive galaxies were undergoing formative mergers at relatively late
times.  

\section{Summary}

We have surveyed an area of 154 sq. arcmin to a 5$\sigma$ 
point-source limit of \Kp\ $=$ 20\fm5, representing the largest 
contiguous field imaged in the near infrared at this depth to date.
From these data, we have selected 8 EROs with \Kp\ $\leq$ 19\fm0 and
with (\R-\Kp) $\geq$ 6\mgn.
 
Six of the 8 bright EROs are resolved at \Kp\, and are, therefore, most
likely to be galaxies at high redshift ($z \geq 0.85$).  These EROs
have a surface density of 0.039 $\pm$ 0.016 arcmin$^{-2}$, or, with
simple assumptions about their range of redshifts, a volume density
comparable to that of Seyfert galaxies.  Interpretations as to the
nature of these galaxies range from old, evolved elliptical galaxies to
young dusty starbursts or AGN.

The remaining 2 EROs are low-mass Galactic stars, either main sequence
M stars or brown dwarfs (Wolf et al. 1998).  Such low-mass stars can be
seen to distances well beyond 100 pc, depending on their intrinsic
luminosity.  Morphological considerations alone rule out the
possibility that the majority of the EROs with \Kp\ $\leq$ 20\mgn\ are
low-mass stars.  The true count of such stars in high-latitude fields
will provide interesting constraints on Galactic structure and the
faint end of the local stellar luminosity function.

EROs appear to be a significant population of galaxies absent from 
samples derived from observations at visual wavelengths.  Future 
work on these objects will distinguish between the old elliptical 
and young dusty starburst interpretations.  Observations at submillimeter 
(ground based) and mid- to far-infrared (SIRTF, NGST) wavelengths will 
be able to quantitatively determine how much dust, if any, is present.  
In either case, the EROs are likely drawn from a population of massive 
galaxies at high redshift.   Whichever interpretation prevails, the EROs 
will provide important clues to the formation and evolution of massive 
galaxies.

\acknowledgments

We are grateful to the team that constructed the OMEGA camera, particularly 
M. McCaughrean and P. Bizenberger, without whom the wide field, deep 
K$^\prime$ survey would not have been possible.  We also thank the staff 
at the Calar Alto observatory for support during the observing runs, and 
MPIA for support of the CADIS key project.  This research was supported 
by the Max-Planck-Society.

\clearpage

\begin{deluxetable}{cccccc}
   \large
   \tablecolumns{6}
   \tablewidth{0pc}
   \tablecaption{ERO properties \label{ERO_data}}
   \tablehead{ \colhead{ID\tablenotemark{a}}   &
               \colhead{Position}              &
               \colhead{\Kp}                   &
               \colhead{\R\tablenotemark{b}}   &
               \colhead{\R-\Kp}                &
               \colhead{FWHM\tablenotemark{c}} \\
               \colhead{}                      &
               \colhead{(J2000.0)}             &
               \colhead{(mag)}                 &
               \colhead{(mag)}                 &
               \colhead{(mag)}                 &
               \colhead{($^{\prime\prime}$)}   }
   \startdata
      C16-ERO1 & 16:24:01.22\ \ $+$55:38:49.0 & 18.56$^{+0.05}_{-0.06}$ & 26.04$^{+1.00}_{-1.00}$ & 7.48 & 2\farcs0 \\
      C16-ERO2 & 16:24:04.99\ \ $+$55:38:26.2 & 18.90$^{+0.08}_{-0.09}$ & 24.97$^{+0.75}_{-0.44}$ & 6.07 & 1\farcs0 \\
      C16-ERO3 & 16:24:28.83\ \ $+$55:43:10.4 & 19.00$^{+0.07}_{-0.07}$ & 25.14$^{+0.33}_{-0.26}$ & 6.14 & 0\farcs4 \\
      C16-ERO4 & 16:24:43.84\ \ $+$55:49:40.6 & 18.98$^{+0.05}_{-0.06}$ & 25.17$^{+1.00}_{-1.00}$ & 6.19 & 1\farcs0 \\
      C16-ERO5 & 16:25:05.53\ \ $+$55:39:54.4 & 18.92$^{+0.07}_{-0.06}$ & 25.64$^{+0.47}_{-0.33}$ & 6.72 & 1\farcs1 \\
      C16-ERO6 & 16:25:13.11\ \ $+$55:48:30.4 & 18.78$^{+0.04}_{-0.05}$ & 24.86$^{+1.00}_{-0.53}$ & 6.08 & 0\farcs9 \\
      C16-LMS1 & 16:25:00.63\ \ $+$55:44:44.3 & 18.57$^{+0.10}_{-0.10}$ & 26.00$^{+1.00}_{-1.00}$ & 7.43 & 0\farcs0 \\
      C16-LMS2 & 16:23:57.24\ \ $+$55:46:27.5 & 16.32$^{+0.05}_{-0.05}$ & 23.06$^{+0.09}_{-0.09}$ & 6.74 & 0\farcs0 \\
   \enddata
   \tablenotetext{a}{ERO:Extremely Red Object; LMS:Low Mass Star (see 
                     section 3.2).}
   \tablenotetext{b}{The uncertainties on the \R-band magnitudes 
                     of objects below the 2$\sigma$ detection limit have 
                     been arbitrarily set to 1\fm0.  This represents taking 
                     a 2$\sigma$ marginal detection to 5$\sigma$, where the 
                     object presumably would have been detected.}
   \tablenotetext{c}{FWHM deconvolved from the seeing.  The two low-mass 
                     stars are unresolved.}
   \normalsize
\end{deluxetable}

\clearpage

\begin{deluxetable}{ccc}
   \tablecolumns{3}
   \tablewidth{0pc}
   \tablecaption{Color Distributions \label{colorhist_data}}
   \tablehead{ \colhead{Magnitude Bin} &
               \colhead{Mean Color}    &
               \colhead{FWHM}          }
   \startdata
      16\fm0 $\leq$ \Kp\ $<$ 17\fm0 & 3.24$\pm$0.06 & 1.34$\pm$0.10 \\
      17\fm0 $\leq$ \Kp\ $<$ 18\fm0 & 3.59$\pm$0.05 & 1.86$\pm$0.08 \\
      18\fm0 $\leq$ \Kp\ $<$ 19\fm0 & 3.73$\pm$0.04 & 2.25$\pm$0.06 \\
      18\fm5 $\leq$ \Kp\ $<$ 19\fm5 & 3.92$\pm$0.04 & 2.60$\pm$0.06 \\
      19\fm0 $\leq$ \Kp\ $<$ 20\fm0 & 3.93$\pm$0.03 & 2.60$\pm$0.04 \\
   \enddata
\end{deluxetable}
 
\clearpage

\end{document}